# Imaging, simulation, and electrostatic control of power dissipation in graphene devices

Myung-Ho Bae[1,2,†], Zhun-Yong Ong[1,3,†], David Estrada[1,2] and Eric Pop[1,2,4,*]

[1]*Micro & Nanotechnology Lab, University of Illinois, Urbana-Champaign, IL 61801, USA*
[2]*Dept. of Electrical & Computer Eng., University of Illinois, Urbana-Champaign, IL 61801, USA*
[3]*Dept. of Physics, University of Illinois, Urbana-Champaign, IL 61801, USA*
[4]*Beckman Institute, University of Illinois, Urbana-Champaign, IL 61801, USA*

[†]These authors contributed equally.
[*]Contact: epop@illinois.edu

We directly image hot spot formation in functioning mono- and bilayer graphene field effect transistors (GFETs) using infrared thermal microscopy. Correlating with an electrical-thermal transport model provides insight into carrier distributions, fields, and GFET power dissipation. The hot spot corresponds to the location of minimum charge density along the GFET; by changing the applied bias this can be shifted between electrodes or held in the middle of the channel in ambipolar transport. Interestingly, the hot spot shape bears the imprint of the density of states in mono- vs. bilayer graphene. More broadly, we find that thermal imaging combined with self-consistent simulation provides a non-invasive approach for more deeply examining transport and energy dissipation in nanoscale devices.

Power dissipation is a key challenge in modern and future electronics.[1,2] Graphene is considered a promising new material in this context, with electrical mobility and thermal conductivity over an order of magnitude greater than silicon.[3,4] Graphene is a two-dimensional crystal of $sp^2$-bonded carbon atoms, whose electronic properties can be tuned with an external gate.[5,6] By varying the gate voltage ($V_G$) with respect to source (S) or drain (D) terminals, as labeled in Fig. 1, the electron and hole densities can be altered, resulting in an ambipolar GFET.[7] At large source-drain voltage bias ($V_{SD}$), the electrostatic potential varies significantly along the channel, leading to an inhomogeneous distribution of carrier types, densities, and drift velocities. The power dissipated is related to the local current density ($J$) and electric field ($F$) in samples larger than the carrier mean free paths ($P' = \mathbf{J} \cdot \mathbf{F}$).[8] Thus, a GFET with large applied bias should have regions of varying power dissipation, tied to the local charge density and electrostatic profile.

Two recent studies[9,10] have revealed the effect of Joule heating in monolayer graphene using Raman thermometry. However, the small size of devices investigated (1-2 µm) did not allow detailed spatial imaging. In this work, we utilize sufficiently large samples (~25 µm) and use infrared (IR) thermal microscopy to observe clear spatial variations of dissipated power, in both monolayer and bilayer graphene devices. In addition, we introduce a comprehensive simulation approach which reveals the coupling of electrostatics, charge transport and thermal effects in GFETs. The combination of thermal imaging and self-consistent modeling also provides a non-invasive method for in situ studies of transport and power dissipation in such devices.

We prepared mono- and bilayer GFETs, as shown in Fig. 1b and described in the Supplementary Information. For consistency, we refer to the ground electrode as the drain and the biased electrode as the source regardless of the majority carrier type or direction of current flow. Sheet resistance vs. gate voltage ($R_S$-$V_{GD-0}$) measurements are shown in Fig. 1c, at low bias ($V_{SD}$ = 20 mV). Here, we subtract the so-called Dirac voltage ($V_0$) which is the gate voltage at the charge neutrality point. Gate voltages lower and higher than $V_0$ provide holes and electrons as the majority carriers, respectively.[11] At low bias the graphene sheet resistance is given by $R_S = 1/[q\mu_0(n+p)]$, where $\mu_0$ is the low-field mobility, $n$ and $p$ are the electron and hole carrier densities per unit area, respectively, and $q$ is the elementary charge. Our new charge density model takes into account thermal generation[12] ($n_{th}$) and residual puddle density[13] ($n_{pd}$) as detailed further below. At high temperatures in our measurements the former often dominates. The fit in Fig.

1c is obtained with $R = R_C + R_S L/W$, where $R_C = 300$ Ω is the measured contact resistance, $L$ and $W$ are the length and width of the GFET. Good agreement is obtained, with only two fitting parameters $\mu_0 = 3590$ cm$^2$V$^{-1}$s$^{-1}$ and $n_{pd} = 1.2 \times 10^{11}$ cm$^{-2}$, consistent with previous reports.[13, 14] We note that at low $V_{SD}$ bias the electrostatic potential and Fermi level are nearly flat along the graphene, and the charge density is constant and determined only by the gate voltage, impurities, and temperature.

On the other hand, a large $V_{SD}$ bias induces a significant spatial variation of the potential in the GFET. This leads to changes in carrier density, electric field, and power dissipation along the channel. In turn, this results in a spatial modulation of the device temperature, as revealed by our IR microscopy. We first consider the monolayer graphene device, as shown in Figs. 1d and 2. The temperature profiles along the graphene channel are displayed in Fig. 1d with various $V_{SD}$ at $V_{GD-0} = -33$ V (strongly hole-doped transport), and the temperature increases linearly with applied power as expected (see Fig. 1d inset).

Figure 2 shows imaged temperature maps with distinct hot spots that vary along the channel with the applied voltage (also see supplementary movie file[15]). This implies that the primary heating mechanism is due to energy loss by carriers within the graphene channel, and not due to contact resistance. However, we note the raw temperature reported by the IR microscope is lower than the actual GFET temperature, and must be corrected before being compared with our simulation results below.[16] Figures 2a-c show raw thermal IR maps of the monolayer GFET for $V_{GD-0}$ = -3.7 V, 3 V, and 12.2 V with $V_{SD}$ = 10 V, 12 V, and 10 V, respectively. These represent three scenarios, i.e. (a) unipolar hole-majority channel, (b) ambipolar conduction, and (c) unipolar electron-majority channel. In the hole-doped regime, at $V_{GD-0}$ = -3.7 V (Figs. 2a,d,g), the hole density is minimum near the drain and a hot spot develops there (left side). As the back-gate voltage increases to $V_{GD-0}$ = 3 V (Figs. 2b,e,h), the graphene becomes electron-doped at the drain. Given that $V_{SD}$ = 12 V, the region near the source remains hole-doped as $V_{GS} = V_{GD}-V_{SD}$ = -9 V. This is an ambipolar conduction mode, with electrons as majority carriers near the drain, and holes near the source as indicated by the block arrows in Fig. 2b. The minimum charge density point is now towards the middle of the channel, with the hot spot correspondingly shifted. At $V_{GD-0}$ = 12.2 V (Figs. 2c,f,i) electrons are majority carriers throughout the graphene channel, and the hot spot forms near the source electrode (right side). In other words, as the gate voltage changes, the device goes from unipolar hole to ambipolar electron-hole and finally unipolar electron conduction, with the hot spot shifting from near the drain to near the source. This is precisely mirrored in the temperature profiles along the graphene channel, as shown in Fig. 2d-f (lower panels).

To obtain a quantitative understanding of this behavior, we introduce a new model of monolayer and bilayer GFETs by self-consistently coupling the current continuity, thermal, and electrostatic (Poisson) equations. This is a drift-diffusion approach[8, 17] suitable here due to the large scale (~25 μm) and elevated temperatures of the GFET, with carrier mean free paths much shorter than other physical dimensions. For example, the electron mean free path may be estimated as[18] $l_n \approx (h/2q)\mu(n/\pi)^{1/2} \approx 30$ nm, for typical $n = 5 \times 10^{11}$ cm$^{-2}$ and $\mu = 3600$ cm$^2$/V·s in our samples. The phonon mean free path has been estimated at[19] $l_{ph} \approx 0.75$ μm in freely suspended graphene, although it is likely to be lower in graphene devices operated a high bias and high temperature on SiO$_2$ substrates. Both figures are much shorter than the device dimensions.

We set up a finite element grid along the GFET, with $x = 0$ at the left electrode edge and $x = L$ at the right electrode. The left electrode is grounded and all voltages are written with respect to it. The electron ($n_x$) and hole ($p_x$) charge densities, velocity ($v_x$), field ($F_x$), potential ($V_x$) and temperature ($T_x$) along the graphene sheet are computed iteratively until a self-consistent solution is found. The "$x$" subscript denotes all quantities are a function of position along the graphene device. We note that the temperature influences[20] the charge density by changing the intrinsic carriers through thermal generation.[12] This is particularly important when the local potential ($V_x$) along the graphene is near the Dirac point, and the carrier density is at a minimum. We also note that both electron and hole components of the charge density are self-consistently taken into account. The model properly "switches" from electron- to hole-majority carriers with the local potential along the graphene, yielding the correct ambipolar behavior of the GFET.

Starting from grid element $x = 0$, the current continuity condition gives:

$$I_D = \text{sgn}(p_x - n_x)qW(p_x + n_x)v_x \tag{1}$$

where the subscript $x$ is the position along the $x$-axis. The carrier densities per unit area are given by $n,p = [\pm n_{cvx} + (n_{cvx}^2 + 4n_{ix}^2)^{1/2}]/2$, where upper (lower) signs correspond to holes (electrons).[21] Here $n_{cvx} = C_{ox}(V_0 - V_{Gx})/q$, $C_{ox} = \epsilon_{ox}/t_{ox}$ is the SiO$_2$ capacitance per unit area, and $V_{Gx} = V_G - V_x$ is the potential difference between the back-gate and graphene channel at position $x$. The intrinsic carrier concentration is[21] $n_{ix}^2 \approx n_{th}^2 + n_{pd}^2$, where $n_{th} = (\pi/6)(k_B T_x/\hbar v_F)^2$ are the thermally excited carriers in monolayer graphene,[12] $n_{pd}$ is the residual puddle concentration,[13] and $T_x$ is the temperature at position $x$. In bilayer graphene, $n_{th} = (2m^*/\pi\hbar^2)k_B T_x \ln(2)$ due to the near-parabolic bands.[22] The velocity ($v_x$) is obtained from the current and charge, and the local field ($F_x$) is calculated from the velocity-field relation[7, 17, 23]

$$F_x = \operatorname{sgn}(p_x - n_x) \frac{v_x}{\mu_0 \left(1 - |v_x/v_{sat}|\right)} \tag{2}$$

which includes the velocity saturation $v_{sat}$ discussed below. The Poisson equation then relates the field to the potential along the graphene[17] as $F_x = \partial V_x/\partial x$. To include temperature we also self-consistently solve the heat equation along the GFET as:

$$A\frac{\partial}{\partial x}\left(k\frac{\partial T}{\partial x}\right) + P'_x - g(T - T_0) = 0, \tag{3}$$

where $P_x' = I_D F_x$ is the Joule heating rate in units of Watts per unit length, $A = WH$ is the graphene cross-section (monolayer "thickness" $H = 0.34$ nm), $k$ is the graphene thermal conductivity, $g$ is thermal conductance to the substrate per unit length, and $T_0$ is the ambient temperature. Interestingly, we note that the device simulations here are quite insensitive to the value of the graphene thermal conductivity ($k \approx 600$–$3000$ Wm$^{-1}$K$^{-1}$),[4, 9, 24] but much more sensitive to the heat sinking path through the SiO$_2$ ($g$) and the exact device electrostatics. Thermal transport in large devices ($L, W \gg$ healing length $L_H \sim 0.2$ μm, see Supplementary Information) is dominated by the thermal resistance of the SiO$_2$ layer, rather than by heat flow along the graphene sheet itself. The thermal transport is reduced to a 1-dimensional problem as in previous work on carbon nanotubes (CNTs).[25, 26] Thus, the thermal coupling between graphene and the silicon backside is replaced by an overall thermal conductance per unit length, $g \approx 1/[L(R_{ox} + R_{Si})] \approx 18$ WK$^{-1}$m$^{-1}$ (see Supplementary Information). This is significantly higher than that of a typical CNT on SiO$_2$ (~0.2 WK$^{-1}$m$^{-1}$),[25, 26] due to the much greater width of the graphene sheet. In addition, heat sinking from CNTs is almost entirely dominated by the CNT-SiO$_2$ interface thermal resistance,[27] whereas thermal sinking from the graphene sheet is primarily limited by the 300 nm thickness of the SiO$_2$ itself.

Figures 2a-c show raw temperature maps taken at the last point in the $I_D$-$V_{SD}$ sweeps from Figs. 2g-i, respectively. Figures 2d-f show actual temperature cross-sections (bottom panels, scattered dots) and simulation results for charge density and temperature (top and bottom panels, lines). Here, the actual temperature of the graphene sheet is obtained based on the raw imaged temperature of Figs. 2a-c (Ref. [16] and Supplementary Information). Field dependence of mobility and velocity saturation are included with an effective mobility $\mu_x = \mu_0(1-|v_x/v_{sat}|)$ in our model. Here, $v_{sat} = v_F|E_{SO}/E_F|$ is the saturation velocity, $v_F \approx 10^6$ m/s is the Fermi velocity, $E_F$ is the Fermi level with respect to the Dirac point (positive for electrons, negative for holes), and $E_{SO} \approx 60$ meV is the dominant surface optical (SO) phonon energy for $SiO_2$.[7] Solid curves from simulations show excellent agreement with the measured $I$-$V$ characteristics (Figs. 2g-i) and good agreement with the measured temperature profiles (Figs. 2d-f)[28] (also see Supplementary Information, Figs. S7-S8). We find that $v_{sat}$ varies from $2.9\times10^7$ cm/s to $8.8\times10^7$ cm/s while the carrier density varies from $3.2\times10^{12}$ cm$^{-2}$ to $3.4\times10^{11}$ cm$^{-2}$.

While the $I$-$V$ characteristics show excellent agreement between experiment and simulation, the temperature profiles provide additional insight into transport and energy dissipation. Best agreement is found near the hot spot locations, marked by arrows in Fig. 2d-f, but a slight discrepancy exists between temperature simulation and data near the metal electrodes. We attribute this in part to inhomogeneous doping and charge transfer on micron-long scales between the metal electrodes and graphene.[29, 30] In addition, recent work has also found that persistent Joule heating can lead to undesired charge storage in the $SiO_2$ near the contacts where the fields are highest,[31] resulting in a possible discrepancy between the experiments and model calculations.

Before moving on, we address a few simulation results which are related to, but not immediately apparent from the temperature measurements. The calculated carrier density profiles along the GFET at each biasing scenario are shown in the upper panels of Figs. 2d-f, respectively. The simulations confirm that temperature hot spots are always located at the position of minimum carrier density along the channel. This occurs near the grounded drain for hole conduction (Fig. 2d) and near the source for electron conduction (Fig. 2f). In ambipolar operation (Fig. 2e) the hot spot forms approximately at $x = -7.5$ μm in both simulation and measured temperature, which is the crossing point of electron and hole concentrations. In this case, the temperature distribution is broader, also in good agreement with the thermal imaging data. Thus, the temperature measurement technique is an indicator of the electron and hole carrier concentrations, as well as the polarity of the graphene device. Combined with our simulation approach, non-invasive IR thermal imaging provides essential insight into the inhomogeneous charge density profile of the GFET channel under high bias conditions. In a sense, this finding is similar to the shift of electroluminescence previously observed in ambipolar carbon nanotubes.[32] However, due to the absence of an energy gap in monolayer graphene, carrier recombination at the pinch-off region results primarily in heat (phonon) dissipation rather than light (photon) emission.

Figure 3 shows the thermal imaging of a bilayer GFET in unipolar hole doped (Fig. 3a with $V_{GD-0}$ = -42 V), ambipolar (Fig. 3b with $V_{GD-0}$ = -12 V), and unipolar electron doped transport regimes (Fig. 3c with $V_{GD-0}$ = 25 V). The qualitative temperature distributions are similar to the respective monolayer GFET cases. For instance, the hot spots in both the hole and electron doped regimes are at the location of minimum carrier density. In ambipolar transport the peak temperature appears near the middle of the bilayer GFET, as shown in Fig. 3b and lower panel of Fig. 3e, similar to the monolayer GFET. However, the temperature profile in bilayer is much broader than in monolayer graphene, a distinct signature of the different band structure and density of states (Fig. 3i vs. Fig. 2i insets). This, in turn, alters the dependence of carrier densities on the electrostatic potential, and the magnitude of the thermally excited carrier concentration $n_{th}$.[22] To take these into account, we include the effective mass $m^* \approx 0.03 m_0$ of the near-parabolic bilayer band structure[33, 34] and the saturation velocity $v_{sat} \approx (E_{OP}/m^*)^{1/2}$ independent of carrier density unlike in monolayer graphene,[23] where $E_{OP} \approx 180$ meV is an average optical phonon energy.[35] The best overall agreement with the bilayer experimental data is found with $\mu_0 = 1440$ cm$^2$V$^{-1}$s$^{-1}$ and $n_{pd} = 0.7 \times 10^{11}$ cm$^{-2}$ as remaining parameters. Using this model all calculated $I_D$-$V_{SD}$ curves (Fig. 3g-i) and temperature distributions of the bilayer GFET (Fig. 3d-f) show excellent agreement with the experimental data. As with the monolayer graphene device, the thermal imaging approach combined with coupled electrical-thermal simulations yields deeper insight into the carrier distributions, polarity, and energy dissipation of the device at high bias. In addition, the agreement between simulations and thermal imaging near the contacts is improved in bilayer graphene, suggesting this system is less sensitive to charge transfer[29, 30] or SiO$_2$ charge storage near the two electrodes.[31]

Before concluding, it is relevant to summarize both fundamental and technological implications of our findings. Of relevance to high-field transport in graphene devices, we found that the power dissipation is uneven, and that the hot spot depends both on device voltages and electrostatics, and the density of states (e.g. monolayer vs. bilayer). The location of the hot spot corresponds to that of minimum charge density in unipolar transport, and to that of charge neutrality in ambipolar operation. Interestingly, the hot spot can be controlled with the choice of voltages applied on the three terminals, such that independent thermal annealing of either the source or the drain, or of any region in between could be achieved, particularly in monolayer graphene.

From a technological perspective, we have shown that graphene-on-insulator (GOI) devices pose similar thermal challenges as those of silicon-on-insulator (SOI) technology.[36-38] For practical applications the $SiO_2$ layer must be thinned to minimize temperature rise, or until parasitic (graphene-to-silicon) capacitance effects limit device performance. Moreover, we have shown that such thermal effects can be modeled self-consistently, by introducing a coupled solution of the continuity, thermal, and electrostatic equations. Finally, the combination of IR imaging and simulations reveals much more than electrical measurements alone, and opens up the possibility of non-invasive thermal imaging as a tool for other studies of high-field transport and energy dissipation in nanoscale devices.

**Supporting Information Available:** Details of sample fabrication and setup, additional model calculations of heat dissipation in graphene, and procedure for obtaining the true graphene temperature from the raw temperature imaged by the infrared scope.

**Acknowledgements:** Fabrication and experiments were carried out in the Frederick Seitz Materials Research Lab (MRL), and the Micro and Nanotechnology Lab (MNTL). We are deeply indebted to R. Pecora and Dr. E. Chow for assistance with the InfraScope setup. We also thank D. Abdula and Prof. M. Shim for help with Raman measurements, Prof. D. Cahill for discussions on IR emissivity, and T. Fang, A. Konar, and Prof. D. Jena for insight on scattering in graphene. The work has been supported by the Nanotechnology Research Initiative (NRI), the Office of Naval Research grant N00014-09-1-0180, and the National Science Foundation grant CCF 08-29907. D. E. acknowledges support from NSF, NDSEG and Micron fellowships.

calculations we find the real graphene temperature rise ($\Delta T$) is proportional to that measured by the IR microscope, and approximately 3 times higher (see Supplementary Information).

$\mu_0$ = 3640 cm$^2$V$^{-1}$s$^{-1}$ ($n_{pd}$ = 1.31×10$^{11}$ cm$^{-2}$), respectively, and this stability is provided by our PMMA passivation (see Supplementary Information, Fig. S8).

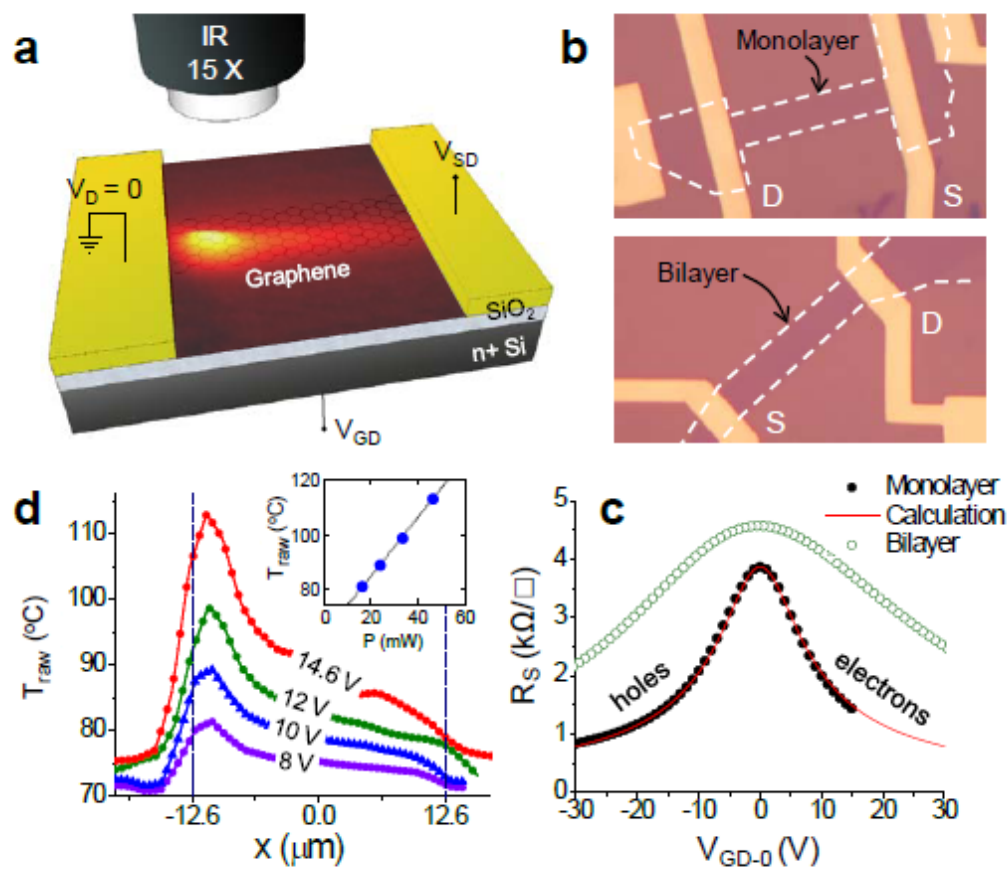

**Figure 1. Graphene field-effect transistors (GFETs). a,** Schematic of GFET and infrared (IR) measurement setup. Rectangular graphene sheet on SiO$_2$ is connected to metal source (S) and drain (D) electrodes. Emitted IR radiation is imaged by 15x objective. **b,** Optical images of monolayer (25.2x6 μm$^2$) and bilayer (28x6 μm$^2$) GFETs. Dashed lines indicate graphene contour. **c,** Sheet resistance vs. back-gate voltage $V_{GD-0} = V_{GD}-V_0$ (centered around Dirac voltage $V_0$) of monolayer (closed points) and bilayer GFETs (open points) at $T_0 = 70$ °C and ambient pressure. **d,** Imaged (raw) temperature along middle of monolayer GFET at varying $V_{SD}$ and $V_{GD-0} = -33$ V (hole-doped regime). Dotted vertical lines indicate electrode edges. The inset shows linear scaling of peak temperature with total power input. Temperature rise here is raw imaged data ($T_{raw}$) rather than actual graphene temperature (see Fig. 2 and Supplementary Information).[16]

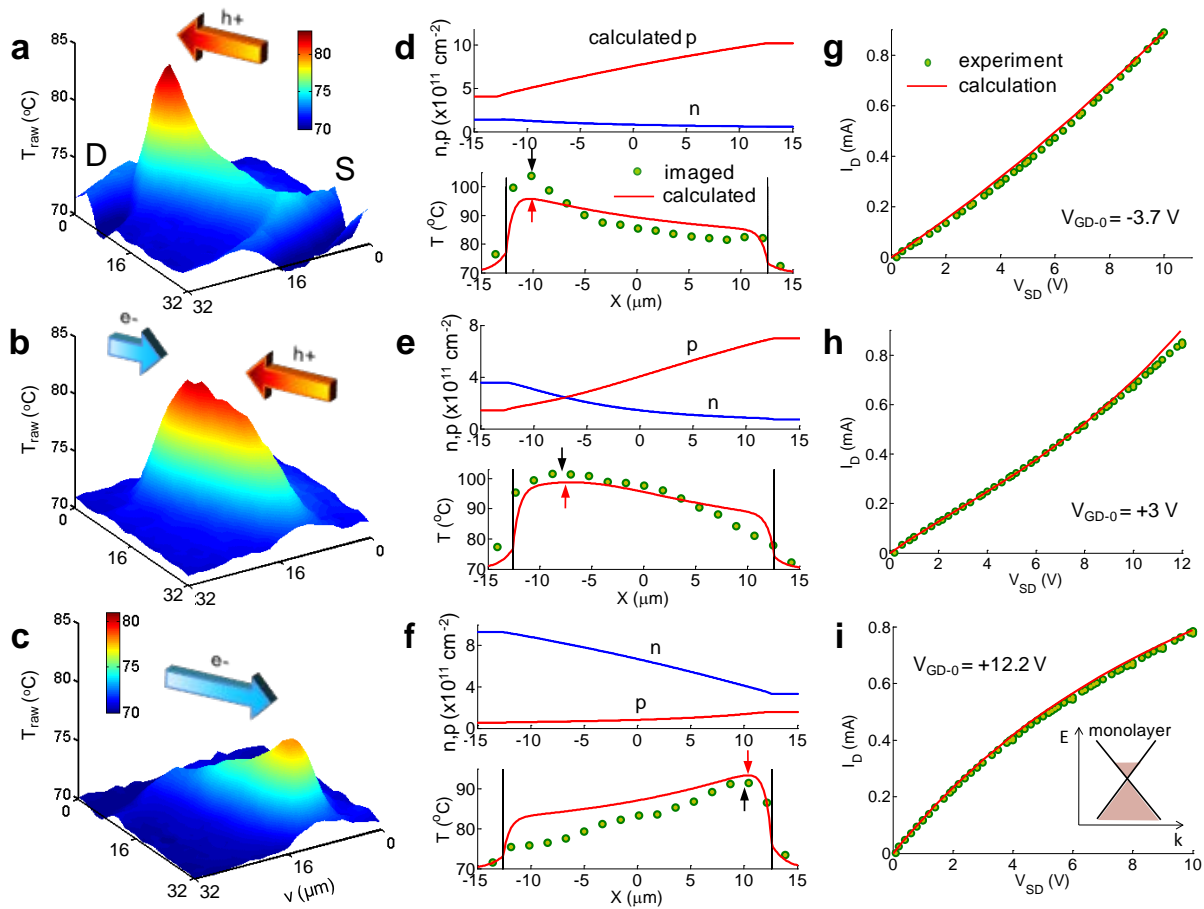

**Figure 2. Electrostatics of monolayer GFET hot spot.** Imaged temperature map at **a,** $V_{GD-0}$ = -3.7 V (hole doped), **b,** 3 V (ambipolar), and **c,** 12.2 V (electron doped conduction) with corresponding $V_{SD}$ = 10 V, 12 V, and 10 V, respectively (approximately same total power dissipation). **d-f,** Charge density (upper panels, simulation) and temperature profiles (lower panels) along the channel, corresponding to the three imaged temperature maps. Symbols are temperature data, solid lines are calculations. Arrows indicate calculated (red) and experimental (black) peak hot spot position, in excellent agreement with each other and consistent with the position of lowest charge density predicted by simulations. **g-i,** Corresponding $I_D$-$V_{SD}$ curves (symbols: experiment, solid lines: calculation). Temperature maps were taken at the last bias point of the $I_D$-$V_{SD}$ sweep.

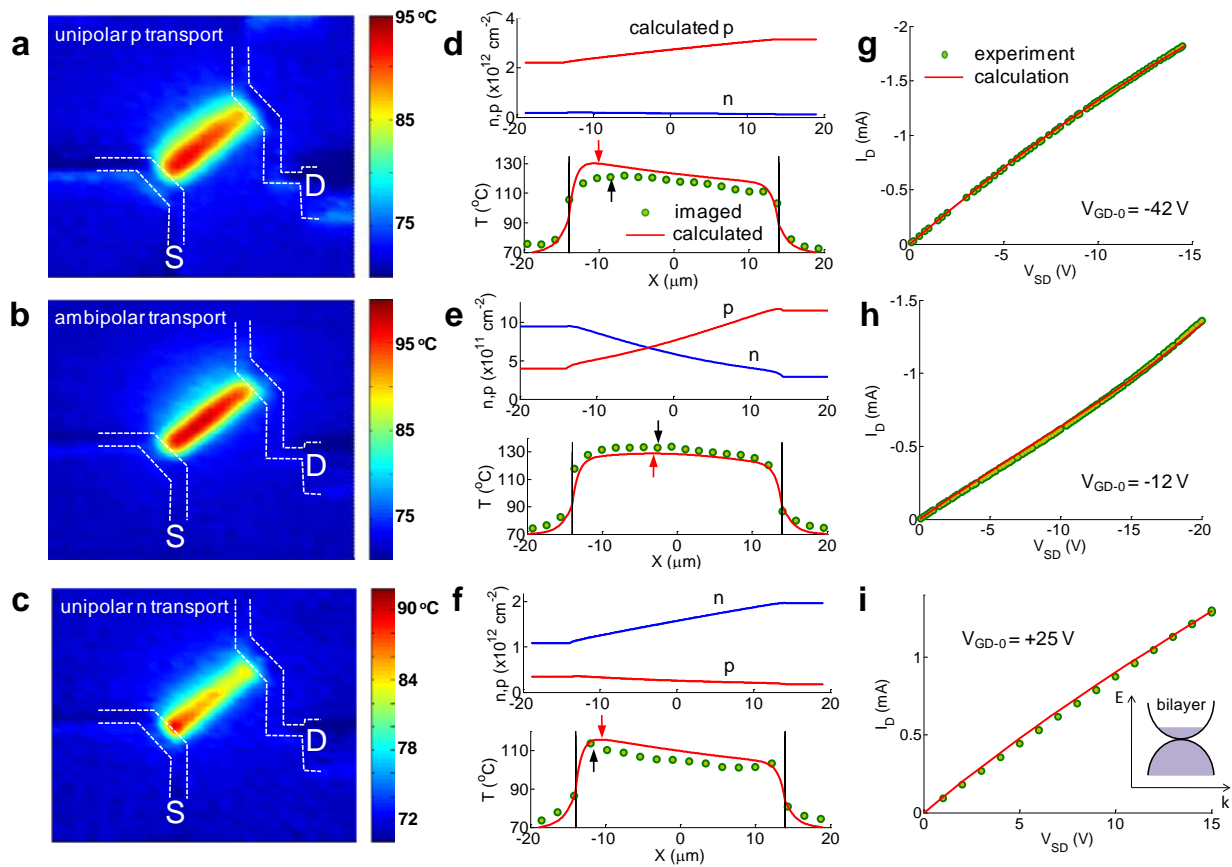

**Figure 3. Electrostatics of bilayer GFET hot spot.** Imaged temperature map of bilayer GFET for **a,** $V_{GD-0}$ = -42 V, **b,** -12 V, and **c,** 25 V with corresponding $V_{SD}$ = -14.5 V, -20 V and 15 V, respectively. **d-f,** Electron and hole density (upper panels, simulation) and temperature profiles (lower panels). Symbols are experimental data, solid lines are calculations. Arrows indicate calculated (red) and experimental (black) hot spot position, in excellent agreement with each other, and with the position of lowest charge density as predicted by simulations. **g-i,** Corresponding *I-V* curves (symbols: experiment, solid lines: calculations). Temperature maps were taken at the last bias point of the *I-V* sweep. The temperature profile of the bilayer GFET is much broader than that of the monolayer (Fig. 2), a direct consequence of the difference in the band structure and density of states (Fig. 2i and Fig. 3i insets).

## Supplementary Information

## Imaging, simulation, and electrostatic control of power dissipation in graphene devices

Myung-Ho Bae[1,2], Zhun-Yong Ong[1,3], David Estrada[1,2] and Eric Pop[1,2,4,*]

[1]*Micro & Nanotechnology Laboratory, University of Illinois, Urbana-Champaign, IL 61801, USA*
[2]*Dept. of Electrical & Computer Engineering, University of Illinois, Urbana-Champaign, IL 61801, USA*
[3]*Dept. of Physics, University of Illinois, Urbana-Champaign, IL 61801, USA*
[4]*Beckman Institute, University of Illinois, Urbana-Champaign, IL 61801, USA*

[*]Contact: epop@illinois.edu

**This section contains:**

1. Sample Fabrication and Experimental Setup
2. Raman Spectroscopy and IR Imaging of GFETs
3. Heat Generation and Dissipation in GFET
4. Additional Figures
5. Supplementary References

**1. Sample Fabrication and Experimental Setup**

We use mechanical exfoliation to deposit graphene onto 300 nm $SiO_2$ with n+ doped ($2.5 \times 10^{19}$ cm$^{-3}$) Si substrate, which also serves as the back-gate.[1] The substrate is annealed in a chemical vapor deposition (CVD) furnace at 400 ºC for 35 minutes in $Ar/H_2$ both before and after graphene deposition.[2] Graphene is located using an optical microscope with respect to markers, confirmed by Raman spectroscopy as shown in Fig. S1,[3] and GFETs are fabricated by electron-beam (e-beam) lithography, as shown in Fig. 1. Electrodes are deposited on the graphene by e-beam evaporation (0.6/20/20 nm Ti/Au/Pd). An additional e-beam lithography step is used to define 6 μm wide graphene channels, followed by an oxygen plasma etch. A 70 nm PMMA (polymethyl methacrylate) layer covers the samples to provide stable electrical characteristics. Electrical and thermal measurements are performed using a Keithley 2612 dual channel source-meter and the QFI InfraScope II infrared (IR) microscope, respectively. IR imaging is performed with the 15× objective which has a spatial resolution of 2.8 μm, pixel size of 1.6 μm, and temperature resolution ~0.1 ºC after calibration.[4] All measurements are made with the IR scope stage temperature at $T_0$ = 70 °C.

**2. Raman Spectroscopy and IR Imaging of GFETs**

*2-A. Raman Spectroscopy*

The difference in the electronic band structure of monolayer and bilayer graphene can be detected by a shift in the Raman spectrum 2D band. Additionally, the 2D band Raman spectra of monolayer and bilayer graphene exhibit a single peak and four peaks respectively. In this study, Raman spectra were ob-tained using a Jobin Yvon LabRam HR 800-Raman spectrometer with a 633 nm laser excitation (power at the object: 3 mW, spot size: 1 μm) and a 100× air objective. Spectra were collected in eight iterations for 16 seconds each. Figure S1 shows the Raman spectra obtained from the GFETs in Fig. 1b, which are monolayer and bilayer GFETs respectively. The Lorentzian fit with the single peak in Fig. S1 gives us a peak frequency of 2643.9 cm$^{-1}$ and a full width at half maximum of 33.6, in agreement with previous findings.[5] In Fig. S1b, a fit result (red curve) for the spectrum of the second sample gives us four relatively shifted peak positions (green curves) with respect to the average frequency of the two main peaks: -56.74, -10.38, 10.38, and 29.71 cm$^{-1}$. These are consistent with previous reports in bilayer graphene.[3]

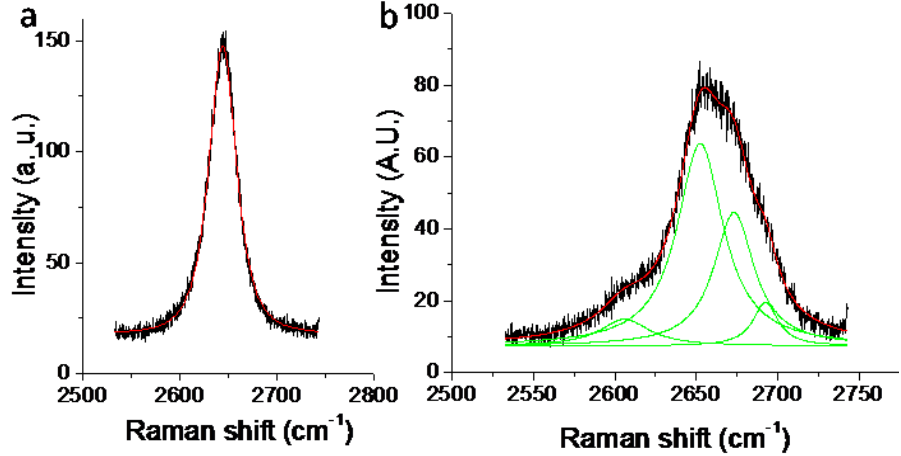

**Figure S1**. 2D band Raman spectra of (**a**) monolayer graphene and (**b**) bilayer graphene at room temperature.

## *2-B. Infrared (IR) Imaging of GFETs with the InfraScope II*

The InfraScope II with a liquid nitrogen-cooled InSb detector provides thermal imaging over the 2–4 μm wavelength range, and working distances of about 1.5 cm with the 15× objective. Thermal mapping with is achieved by sequentially capturing images under different bias conditions. Therefore, the sample is mechanically fixed to the stage to prevent movement during measurements. The InfraScope sensitivity improves with increasing base temperature ($T_0$) of the stage because the number of photons emitted increase as $T_0^3$. However, high temperatures can create convection air currents, resulting in a waved image. Therefore, the recommended stage temperature is between 70 and 90 °C.[4]

Before thermal mapping the GFET, the sample radiance is acquired at the base temperature with no applied voltage ($V_{GD} = V_{SD} = 0$ V). The radiance image is used to calculate the emissivity of the sample at each pixel location before increasing the $V_{SD}$ bias. Figure S2 shows the emissivity image of (a) monolayer and (b) bilayer GFETs, where light blue colored regions indicate electrodes. After acquiring a radiance reference image, an unpowered temperature image is acquired to confirm the set-up as shown in Fig. S3a, where the temperature error is approximately ±0.5 °C. With these pre-conditions, we took thermal images under various applied voltages (Fig. S3b).

The emissivity of the metal electrodes must be considered in order to resolve their temperature. For example, since the emissivity of polished Au is ~0.02 between $T = 38$-260 °C, QFI recommends a background stage temperature between 80 and 90 °C.[4,6] In our experiment, we used electrodes with Pd (20 nm) on top of an Au layer (20 nm) to increase the resolution of the instrument over the contacts (the emissivity of Pd is ~0.17 between $T = 93$-399 °C).[6]

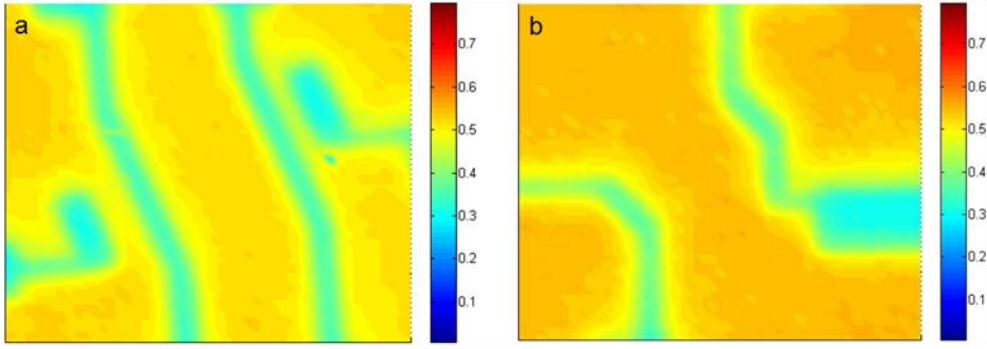

**Figure S2.** Emissivity image of **(a)** monolayer graphene and **(b)** bilayer graphene.

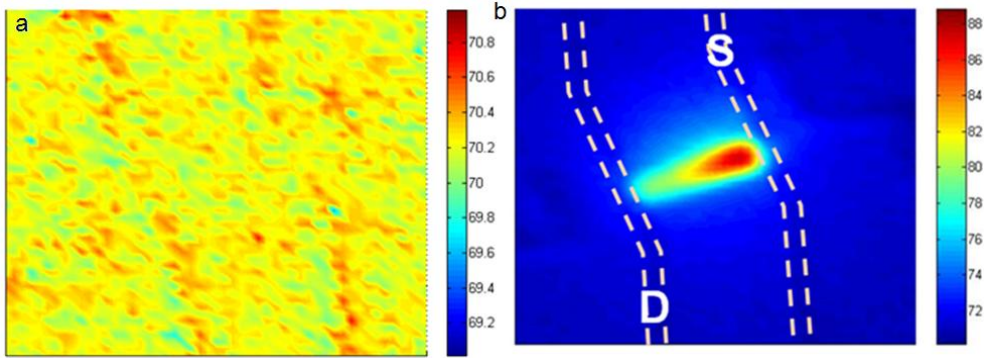

**Figure S3.** IR microscopy image of monolayer GFET **(a)** without applied voltage and **(b)** with $V_{SD}$ = -12 V and $V_{GD-0}$ = -28 V at base $T_0$ = 70 °C, where the region taken in (a) is the same region with Fig. S2a (note different scale bars).

### 3. Heat Generation and Dissipation in GFET

In our simulation code, the temperature profile along the graphene channel is obtained numerically, using the uneven heat generation profile from the electrical transport (described in the main body of the manuscript). However, additional physical insight can be obtained if we consider a simpler scenario of uniform heat generation $Q$ and long fin (longer than carrier scattering lengths) such that ballistic effects may be neglected. In this case, the temperature profile along the graphene can be understood with the simpler one-dimensional fin equation:[7]

$$\frac{d^2T}{dx^2} - \frac{T-T_0}{L_H^2} + \frac{Q}{k} = 0$$

Given the geometry of the device, this suggests the temperature distribution has a characteristic spatial ("healing") length $L_H = (t_{ox}t_G k_G/k_{ox})^{1/2} \approx 0.2$ μm, where $t_G \approx 0.34$ nm is the graphene thickness and $k_G \approx 600$ Wm$^{-1}$K$^{-1}$ is the graphene thermal conductivity on SiO$_2$.[8] The healing length is a measure of the lateral temperature diffusion from a heat source along the graphene. The small $L_H$ means the local heat generation in the graphene is minimally diffused laterally, and is smaller than our IR scope resolution. In other words, there is little lateral broadening of the hot spot, and the heat flow path is mostly directed downwards through the 300 nm SiO$_2$ layer. Thus, the temperature profile of the graphene qualitatively represents the heat generation profile.

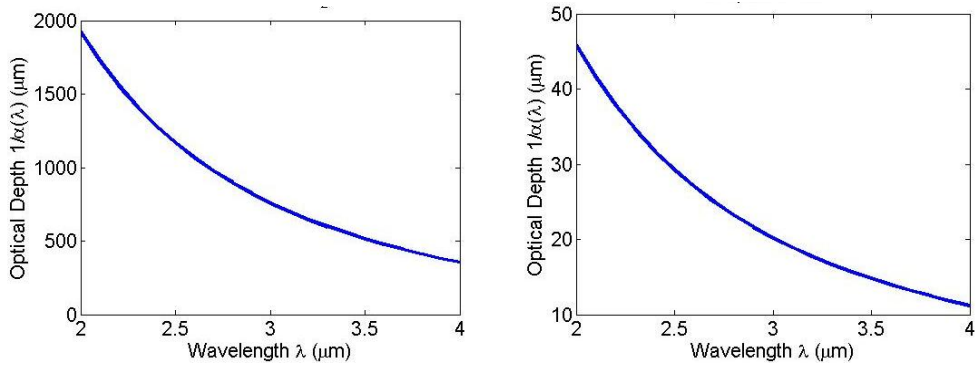

**Figure S4.** Wavelength dependence of the optical depth of (left) thermally grown $SiO_2$ calculated using the Lorentz-Drude oscillator model and (right) heavily doped Si using the Free Carrier Absorption theory.

### 3-A. Infrared Properties of PMMA, $SiO_2$ and Si Layers

Our devices are covered with a ~70 nm layer of PMMA to prevent spurious sample oxidation and significant shift in Dirac voltage ($V_0$) after repeated measurements. The transmittance of PMMA in the infrared has been previously measured and is ~90% for 800 nm thick films in the 2-4 μm wavelengths.[9] Thus, our thinner PMMA films are >99% transparent over our thermal IR imaging range.

To determine the near-infrared optical properties of the thermally grown $SiO_2$ layer and the Si substrate we calculated the wavelength-dependent absorption coefficient of thermally grown $SiO_2$ from the Lorentz-Drude oscillator model[10] of its near-IR dielectric function. The absorption coefficient is given by $\alpha(\lambda) = 4\pi n_I/\lambda$, where $\lambda$ is the wavelength and $n_I$ the imaginary part of the complex refractive index. We also calculated the wavelength-dependent absorption coefficient for doped silicon using the free carrier absorption theory.[11] The measured input parameters for the carrier density and resistivity of the doped silicon are $2.5 \times 10^{19}$ $cm^{-3}$ and $2.7 \times 10^{-3}$ $\Omega \cdot cm$ respectively. The optical depth for $SiO_2$ and Si is given by $1/\alpha(\lambda)$ and is shown in the plots of Fig. S4.

Since the optical depth for $SiO_2$ of near-IR radiation in the region greatly exceeds the thickness of the $SiO_2$ layer (300 nm), we can assume that the $SiO_2$ is effectively transparent. The transparency of $SiO_2$ in this region has been confirmed experimentally by others.[12] On the other hand, we find that the optical depth for doped Si is much smaller and is of the order of ~10 μm, since the emission spectrum over the 2-4 μm range is heavily weighted toward the longer wavelengths. Moreover, the temperature in the Si is highest near the Si-$SiO_2$ interface, strongly weighing the number of IR imaged photons. Hence, we can assume that the IR Scope is effectively reading a thermal signal corresponding to a combination of the graphene temperature and that of the substrate near the Si-$SiO_2$ interface (see sections 3-B & 3-C below).

### 3-B. Finite Element Modeling of Heat Spreading in Substrate

In order to relate the imaged temperature with the actual temperature of the graphene transistor, we consider the calculations and schematic in Fig. S5. The thermal resistance of the $SiO_2$ can be written as $R_{ox} = t_{ox}/(k_{ox}WL) \approx 1417$ K/W underneath the monolayer GFET, where $k_{ox} \approx 1.4$ $Wm^{-1}K^{-1}$ is the thermal conductivity of $SiO_2$ in this temperature range.[13] The thermal boundary resistance between graphene and $SiO_2$ has recently been estimated[14] at ~$10^{-8}$ $m^2K/W$, however this is a relatively small contribution (66 K/W or ~5%) compared to that from the 300 nm $SiO_2$ below the graphene, and from the silicon wafer.

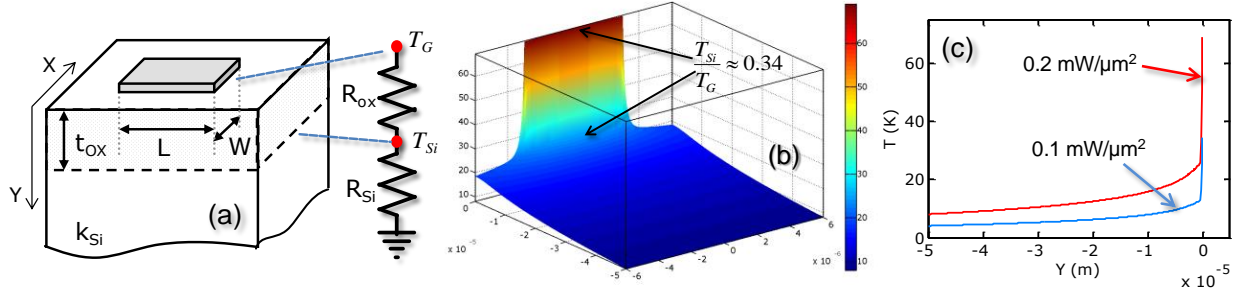

**Figure S5.** Modeling heat dissipation from graphene on SiO$_2$. (a) Schematic of graphene on $t_{ox}$ = 300 nm SiO$_2$. The thermal resistance of the oxide ($R_{ox}$) and that of the silicon substrate ($R_{Si}$) are given in the text. (b) Finite-element simulation of temperature drop across the oxide and silicon, at 0.2 mW/μm$^2$ graphene power density. $T_{Si}$ and $T_G$ represent the temperature *rise* at the graphene and silicon surface, with respect to the silicon backside. (c) Cross-section of temperature through the oxide and silicon substrate, at two different graphene power inputs.

At the same time, the SiO$_2$ film is very thin with respect to the lateral extent of the large ($W \times L = 6 \times 25.2$ μm$^2$) monolayer graphene device, suggesting insignificant lateral heat spreading within the oxide. Thus, the "thermal footprint" of the graphene at the Si/SiO$_2$ surface is still, to a very good approximation, equal to $6 \times 25.2$ μm$^2$. This allows us to write another simple model for the thermal spreading resistance into the silicon wafer, $R_{Si} \approx 1/[2k_{Si}(WL)^{1/2}] \approx 813$ K/W, where $k_{Si} \approx 50$ Wm$^{-1}$K$^{-1}$ is the thermal conductivity of the highly doped substrate above 70 °C temperature range.[15] The ratio between the temperature rise of the graphene and that of the silicon surface can be estimated with the thermal resistance circuit shown in Fig. S5a as $T_G/T_{Si} = 1 + R_{ox}/R_{Si} \approx 2.9$. A similar result is obtained and confirmed via finite element (FE) modeling of the heat spreading beneath the graphene sheet. A typical result is shown in Fig. S5b, and vertical temperature cross-sections through the graphene, SiO$_2$ and silicon are shown in Fig. S5c. The ratio between the temperature of the graphene and that of the Si/SiO$_2$ interface is once again found to be approximately 3:1, for graphene sheets of our dimensions, on 300 nm SiO$_2$ thickness.

### 3-C. Real Temperature of Graphene Sheet

When thermal imaging of the graphene (monolayer or bilayer) and the silicon substrate are initially calibrated at the same temperature ($T_G = T_{Si}$), the power or radiance over the InfraScope wavelength range (2–4 μm) is the sum of the radiance from the graphene (G) and the silicon substrate (Si), given by $P_{tot}(T) = P_G(T) + P_{Si}(T)$. In general, the radiance is the integral of the emitted power per unit wavelength from 2 to 4 μm. Hence, the surface temperature as measured by the InfraScope is a function of the radiance i.e. $T(P_{tot})$. When the graphene is at the same temperature as the silicon, as during calibration, $P_G$ can be neglected because its emissivity ($\epsilon_G \approx 0.023$ for monolayer and 0.046 for bilayer) is much smaller than that of silicon ($\epsilon_{Si} \approx 0.6$, as obtained directly from the InfraScope). Hence, the emissivity as measured by the InfraScope is that of silicon at the same temperature.

However, when the temperature of the graphene increases during Joule heating ($T_G > T_{Si}$), the radiation power from the graphene begins to contribute to the detected power in the InfraScope as shown in Figs. S6a (monolayer graphene) and b (bilayer graphene). But, the InfraScope still measures a single surface temperature $T$ based on the total power emitted by the graphene and the Si surfaces with a single calibrated emissivity (of Si) (see Figs. S6c and d). When the GFET undergoes Joule heating, we estimate the graphene temperature rise to be roughly $R_{\Delta T} \sim 2.9$ times the temperature rise in silicon, as discussed in Section 3-B above. Thus, $T_G = T_{Stage} + \Delta T_G = T_{Stage} + R_{\Delta T} \Delta T_{Si}$ and $T_{Si} = T_{Stage} + \Delta T_{Si}$ where $T_{Stage}$ is the stage temperature and $\Delta T_G$ and $\Delta T_{Si}$ are the temperature rise in the graphene and in the silicon, respectively. Therefore, the total radiance over the detectable wavelength range is given by $P_{tot}(T_{Stage}+\Delta T_{IR}) = P_G(T_{Stage} + R_{\Delta T} \Delta T_{Si}) + P_{Si}(T_{Stage} + \Delta T_{Si}) = f(\Delta T_G)$ where $\Delta T_{IR}$ is the temperature rise measured by the InfraScope i.e. the total radiance is a function of $\Delta T_{Si}$ and thus, a bijective function of $\Delta T_G$. However, the relationship between $\Delta T_{IR}$ and $\Delta T_G$ does not lend itself to a closed form. So, in practice, $\Delta T_G$ as a function of $\Delta T_{IR}$ is determined by first computing the radiance for a given $\Delta T_{IR}$ and then finding the corresponding $\Delta T_{Si}$ and $\Delta T_G$ for that computed radiance. In other words, $\Delta T_G = f^{-1}(P_{tot}(T_{Stage}+\Delta T_{IR}))$.

Since the InfraScope still uses the Si emissivity to get $\Delta T_{IR}$ based on $P_{tot}$, $\Delta T_{IR}$ is always $> \Delta T_{Si}$. This occurs because the graphene introduces a contribution to the total radiance measured by the InfraScope when $T_G > T_{Si}$. As shown in Fig. S6e-f, we work backwards as explained earlier, and numerically convert the measured temperature to the actual temperature in the graphene based on the Planck radiation law accounting for the different emissivities of three materials (monolayer, bilayer graphene, and Si), and the geometrical factors explained in Section 3-B.

## 4. Additional Figures

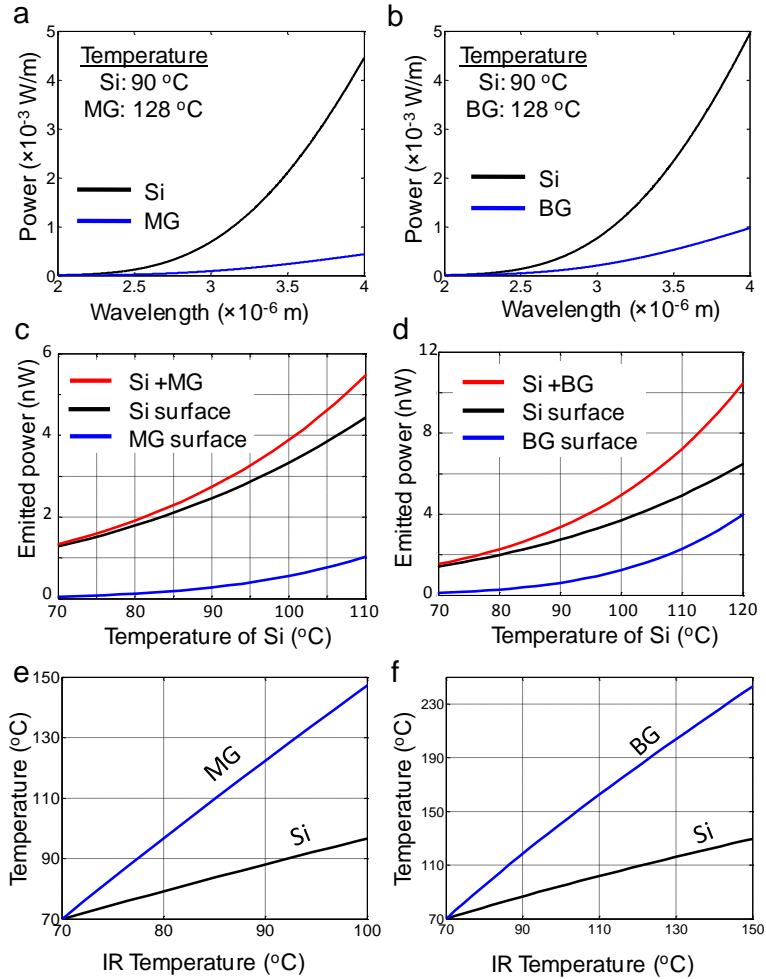

**Figure S6**. Radiation power density as a function of relevant IR wavelength **(a)** from Si and monolayer graphene (MG) surfaces (area: 6×25.2 µm²) and **(b)** from Si and bilayer graphene (BG) surfaces (area: 6×28 µm²) at given temperatures. **(c)-(d)** Total emitted power vs. temperature of Si surface integrated over the wavelength range 2-4 µm from Si, graphene and their combination, where temperatures of graphene are obtained by $2.9(T_{si}-70\ ^oC)+70\ ^oC$. **(e)-(f)** Correspondence between real temperature of graphene and Si surface vs. temperature read by the IR scope.

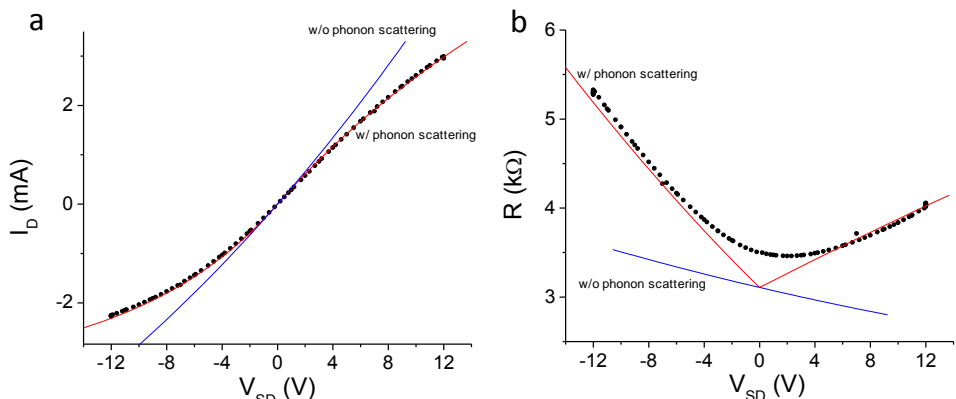

**Figure S7. (a)** $I_D$-$V_{SD}$ curve (scattered points) at $V_{GD\text{-}0}$ = -33 V (a highly hole doped region) of monolayer GFET, which is fitted by two cases: without phonon scattering (blue curve) and with phonon scattering (red curve). **(b)** $R$-$V_{SD}$ curve (scattered points) corresponding to (a), where blue and red curves are fit result without and with phonon scattering, respectively. Here, we used $\mu_0$ = 3780 cm$^2$V$^{-1}$s$^{-1}$ and $n_{pd}$ =1.15×10$^{11}$ cm$^{-2}$ V to fit the $R$-$V_{GD\text{-}0}$ curve for the calculations.

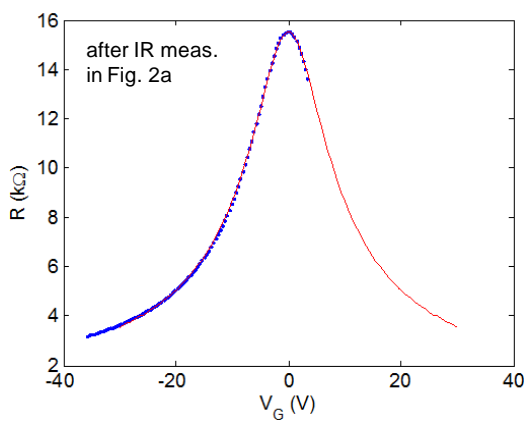

**Figure S8.** Measured $R$-$V_{GD\text{-}0}$ **(a)** after high-current annealing and collecting IR data from Fig. 2a, where scattered points are experimental data and solid curves are fit results. Numerical fitting to measured $R$-$V_{GD\text{-}0}$ curves give $\mu_0$ = 3500 cm$^2$V$^{-1}$s$^{-1}$ ($n_{pd}$ =1.45×10$^{11}$ cm$^{-2}$ V), showing that repeated thermal cycling did not significantly affect the sample properties.

## 5. Supplementary References